# Time-delayed coupled chaotic semiconductor lasers show new types of synchronization: Experiment and theory

Y. Aviad[1], I. Reidler[1], M. Zigzag[2], M. Rosenbluh[1] and I. Kanter[2*]

[1] Department of Physics, The Jack and Pearl Resnick Institute for Advanced Technology, Bar-Ilan University, Ramat-Gan, 52900 Israel

[2] Department of Physics, Bar-Ilan University, Ramat-Gan 52900, Israel

*ido.kanter@biu.ac.il

PACS: 42.65.Sf, 05.45.Xt

Topologies of two, three and four time-delay-coupled chaotic semiconductor lasers are experimentally and theoretically found to show new types of synchronization. Shifted zero-lag synchronization is observed for two lasers separated by long distances even when their self-feedback delays are not equal. Shifted sub-lattice synchronization is observed for quadrilateral geometries while the equilateral triangle is zero-lag synchronized. Generalized zero-lag synchronization, without the limitation of precisely matched delays, opens possibilities for advanced multi-user communication protocols.

In spite of the extreme sensitivity of chaotic orbits to small perturbations, it has recently been demonstrated theoretically and experimentally that a network of mutually time-delay coupled identical chaotic units can synchronize. Generally there are two classes of synchronized states; zero-lag synchronization (ZLS) where all units are synchronized with zero phase shift and sub-lattice synchronization where the network units are split into two clusters, with each cluster in ZLS but with no synchronization between clusters [1-3].

Chaotic semiconductor (SC) lasers have recently been shown to be an extremely useful system for the generation of broadband physical random numbers [4-5]. In addition, synchronization of networks of coupled chaotic lasers, separated by distances much greater than the individual laser coherence length, is the basis of several recently proposed broadband secure communication in private and public channel schemes [6-8]. There are two fundamental issues, however, that need to be resolved before such lasers can be used in a communication network. ZLS and sub-lattice synchronization have to be experimentally demonstrated for small networks configured in various geometries, where the chaos emerges due to optical feedback, as opposed to the recently demonstration emergence of chaos via external optoelectronic feedback [9-11]. The second issue is the constraining limitation of having to use precise and specific ratios among delay times of the network [12-13]. In this Letter we demonstrate various generalized types of synchronization for classes of geometric configurations in small networks that eliminate precise delay-time requirements..

In our most basic small network block, two Fabry-Perot semiconductor (SC) diode lasers, $LD_A$ and $LD_B$, operating near 656 nm, are mutually coupled as well as possessing self feedback as shown in Fig. 1(a). Signals from each of the lasers are coupled using a 50/50 beam splitter (BS), where $\tau_a$ (green) and $\tau_b$ (red) denote the optical delay times to the BS from $LD_A$ and $LD_B$, respectively. The combined signal is coupled into an optical fiber which extends the optical delay by $\tau$=1005 ns (brown) and is retro reflected by mirror, M. A BS near each laser samples a small portion of the beam into a fast (50 GHz bandwidth) photodetector. The dc output, measured via a 40 GHz bandwidth bias-T, measures the average dc power of the lasers, while the ac currents are digitized by a 12 GHz bandwidth, 40 GS/s digital oscilloscope (Tektronix TDS 6124C). Two natural density (ND) filters, consisting of a half-wave plate and two polarizing BSs are used to compensate unavoidable deviations from ideal 50/50 splitting and to ensure that each laser receives the same input intensity. The two lasers are temperature tuned to nearly identical wavelengths while their injection current to threshold current ratios are maintained nearly equal at p= $I/I_{th}$=1.2. When subjected to feedback, the lasers display chaotic behavior, consisting of very short and random spiking of the laser intensity [14].

The experimental setup is depicted schematically in Fig. 1(b), where the mutual coupling delay time is $T_M=2\tau+\tau_a+\tau_b$ and self-feedback delay times are $T_A=2(\tau+\tau_a)$ and $T_B=2(\tau+\tau_b)$. Such a geometry obeys the simplest necessary sum rule for synchronization, where the sum of the two self-feedback delay times is equal to twice the mutual delay time [15]

$$T_A + T_B = 2T_M \quad (1)$$

while $\tau_a$, $\tau_b$ and $\tau$ can be independently varied. The intensity correlation between the two lasers is calculated from matching, 10 ns long time segments from each detector, averaged over all segments within an observation time of 4 μs. The sampling time and step size in the cross correlation calculation was 25 ps. Low frequency fluctuations (LFFs) [16-17] which result in temporal synchronization breakdowns, occur on time scales greater than the overall feedback time, which in this case was ~ 2 µs. We could thus choose data streams which did not contain LFFs.

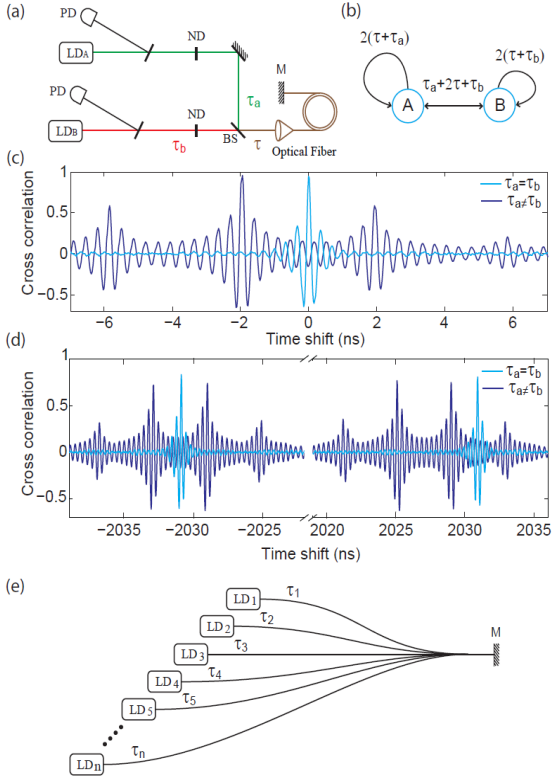

FIG. 1 (color online). (a) Experimental setup: two mutually coupled SC lasers, $LD_A$ and $LD_B$. PD, photodiode; M, mirror. $\tau_a$ and $\tau_b$ are the optical delay times between $LD_A$ and $LD_B$ and the coupling beam splitter, BS, respectively. $\tau$ indicates the optical delay between the BS and mirror, M. (b) Schematic diagram of the experimental configuration. (c) Shifted cross correlation between the chaotic intensities of $LD_A$ and $LD_B$. ZLS for identical self feedback and mutual coupling delays, $\tau_a=\tau_b$ (light blue), and shifted ZLS for $\tau_a \neq \tau_b$ (dark blue) (d) Shifted cross correlation around the mutual delay time between the two lasers. (e) A generalized star configuration of n SC lasers. The output intensity of the lasers with unequal optical delay times, $\{\tau_k\}$, is combined with equal weight and reflected by a passive mirror.

By tuning the delay times between the lasers and coupling BS to be equal, $\tau_a=\tau_b$, ZLS with cross correlation, $\rho=96$, is achieved, Fig. 1(c) (light blue). The correlation repeaks at integer multiples of the round trip time, $2(2\tau+\tau_a+\tau_b)$, the first pair of which are shown by the light blue trace in Fig. 1(d).

Theoretical analysis of our experimental configuration under the constraint of Eq. (1), for *any distance* of the two lasers to the BS, predicts a shifted ZLS correlation peak at $\Delta\tau = \tau_b-\tau_a$ [15]. In other configurations, where the constraint of Eq. (1) is fulfilled by tuning all three delays, it is difficult to observe a shifted ZLS because each detuning has to be precise with a precision approximately equal to the coherence length of the lasers [12-13], as well as the constraint of equal feedback intensity to the two lasers [1, 14]. By measuring the cross correlation of the lasers when the self-feedback is blocked, we determine $\tau_a-\tau_b\sim-1.95$ ns. Indeed, when the two lasers are coupled as in Fig. 1(a), the cross correlation between the two lasers revealed a shifted ZLS, $\rho=95$, at $\Delta\tau\sim-1.95$ ns, Fig. 1(c) (blue). The notable oscillations with period ~0.3 ns in the cross correlation are attributed to the relaxation oscillation frequency of the SC lasers. Similar oscillations were also observed in simulations of the Lang-Kobayashi equations [18], especially at low coupling intensities. In the experiment the feedback is weak, due to various losses in the optics and the fiber, and this leads to the strong relaxation oscillation shown in Fig. 1(c).

A laser with self-feedback is quasiperiodic in its intensity fluctuations with a period equal to the feedback delay time. For lasers A and B this occurs at times $T_{A,n}=2(\tau+\tau_a)n$ and $T_{B,n}=2(\tau+\tau_b)n$, where n is an integer. Because the two lasers are synchronized, both periodicities must be present in the common synchronized fluctuations of the lasers. As a consequence the correlation has additional peaks at time shifts $\pm T_{A,n}$ and $\pm T_{B,n}$, with respect to the shifted ZLS peak (Fig. 1(d)). The quasiperiodicity at these two frequencies produces sidebands at $\pm(T_{A,n}-T_{B,n})=\pm 2(\tau_a-\tau_b)$, which are observed surrounding both the shifted ZLS peak (Fig. 1(c)) and the recurring correlations at $\pm T_{A,1}$ and $\pm T_{B,1}$ (Fig. 1(d)). Twice the mutual feedback repetition time is precisely equal to the sum of the two self-feedback times, as in Eq. (1), which insures that all correlation peaks are at the same time shifts.

The experimental scheme for shifted ZLS, without requiring precise matching of self-feedback and mutual delays, can be generalized to multiple user synchrony. A generalized star configuration [19] of n SC lasers is shown in Fig. 1(e), where the output intensity of all the lasers is combined and mixed with equal weight and upon reflection is redistributed back to the entire network but with unequal optical delay times, $\{\tau_k\}$. Fig. 1(a-b) is a special case of this geometry with n=2. Each pair (m,k) in the general configuration thus has shifted ZLS at a time shift corresponding to the delay time difference between the pair, $\tau_m-\tau_k$. The solutions of the time shifts for all possible pair combinations are consistent with each other even when multiple traverses are allowed. For instance, the time shift for pair (m,k), $\tau_k-\tau_m$, is identical to the sum of all possible delays in getting from m to k;. (m,m+1), (m+1,m+2), …, (k-1,k)

$$\sum_{i=m}^{k-1} \tau_{i+1} - \tau_i = \tau_k - \tau_m \quad (2)$$

This multiple user synchrony was confirmed in simulations of the Lang-Kobayashi equations [18] up to n=7 and might be a source for abundant multiple-user communication and secure communication protocols.

A modification of the geometry of Fig. 1(a), which couples four lasers but without self-feedback is shown in Figure 2(a), where the mirror, M, is replaced by two additional SC lasers, $LD_C$ and $LD_D$, with optical delay to their coupling BS, $\tau_c$ and $\tau_d$, respectively. The delay time $\tau$ between the two coupling BSs is arbitrary and can be made as long as necessary by insertion of a fiber, though in the experiment described below it was maintained at a free space value of a few ns. The resulting system has quadrilateral geometry and is shown schematically in Fig. 2(b). It is important to note that the four independent delays $(\tau_a, \tau_b, \tau_c, \tau_d)$ in this geometry do not form an unconstrained quadrilateral, since the difference between the delays (D,A)-(A,C) has to be equal to (D,B)-(B,C).

For the limiting case, $\tau_a=\tau_b=\tau_c=\tau_d$, the geometry reduces to a square, for which sub-lattice synchronization was theoretically predicted, with the four lasers split into diagonal pairs, (A,B) and (C,D), which are ZLS [20], while the correlation between lasers belonging to adjacent pairs (A,C; C,B; B,D; D,A) is characterized by attenuated correlations at multiples of the delay time between the lasers, $\pm 2\tau_a+\tau$, similar to a face-to-face configuration of two lasers.

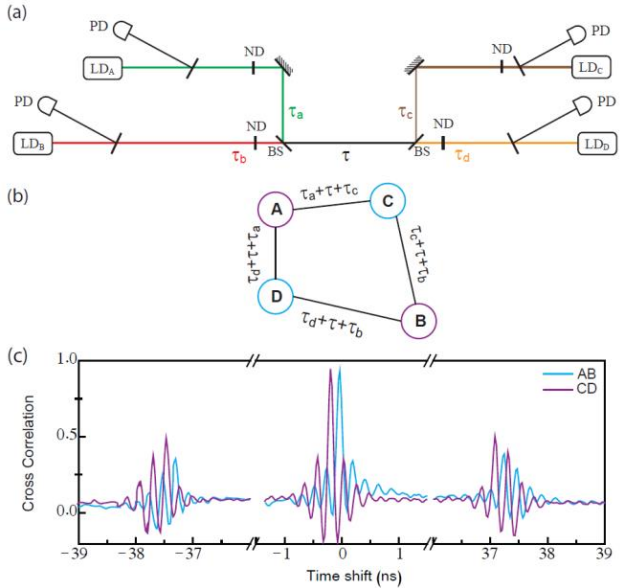

FIG. 2 (color online). (a) Experimental setup of four SC lasers in a quadrilateral geometry. The optical time delays between lasers A and B (C and D) and their coupling beam splitter is $\tau_a$ and $\tau_b$ ($\tau_c$ and $\tau_d$), respectively. $\tau$ is the the optical delay between the two coupling BSs. (b) Schematic diagram of the quadrilateral experimental setup. (c) Shifted cross correlation of the chaotic intensities between lasers A and B (light blue) and C and D (purple), indicating shifted ZLS between diagonal pairs.

Here we experimentally observe generalized sub-lattice synchronization for the general quadrilateral geometry. Figure 2(c) shows the measured shifted cross correlation, with $\rho$=93 between $LD_A$ and $LD_B$ at a shifted time $\tau_a-\tau_b \sim$ -0.03 ns and $\rho$=94 between $LD_C$ and $LD_D$ at a shifted time of $\tau_c-\tau_d \sim$ -0.195 ns. Additional attenuated correlation peaks between the diagonal lasers pairs occur at ±37.28 ns around the central shifted ZLS peaks. The adjacent pairs, however, have no correlation near zero time delay but have correlation (not shown) at times corresponding to an effective face to face delay time as described below. Note that the background correlation in Fig. 1(c) is slightly above zero because for these experimental parameters LFFs occur at times > ~100 ns and in order to avoid data segments containing LFFs only the highest 50% of the correlation segments are averaged.

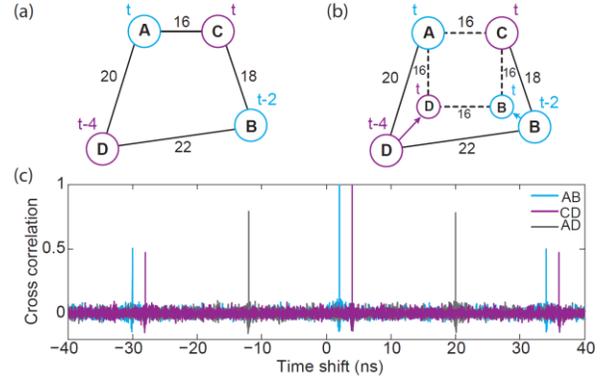

FIG. 3 (color online). (a) Configuration for numerical simulation with $\tau_a$=2 ns, $\tau_b$=4 ns, $\tau_c$=3 ns, $\tau_d$=7 ns and $\tau$=11 ns. Pairs of lasers with the same color (light blue and purple) are shifted ZLS as indicated by the relative time in ns units. (b) The transformation from a quadrilateral geometry to a square. Dashed lines represent the effective time delays after compensating for the synchronization time shift. (c) Calculated shifted cross correlation of the chaotic intensities of the quadrilateral geometry, Fig. 3(a).

In order to understand the emergence of the generalized sub-lattice synchronization we performed simulations using the Lang-Kobayashi equations where the explicit equations and their parameters are given in [16, 18] and $p=I/I_{th}$=1.2, similar to the experimental operating parameters. The time delays for the chosen quadrilateral geometry were arbitrarily selected to be $\tau_a$=2 ns, $\tau_b$=4 ns, $\tau_c$=3 ns, $\tau_d$=7 ns and $\tau$=11 ns, thus yielding a ZLS delay between laser A and B, $\tau_a-\tau_b$=2 ns and between lasers C and D, $\tau_c-\tau_d$=4 ns, Fig. 3(a). To achieve a better understanding of this type of generalized synchronization, the quadrilateral geometry is transformed into a square, Fig. 3(b), in the following way: Since we know that laser D has a shifted ZLS with laser C of 4 ns we move along the laser emission line of laser D by 4 ns towards A and B, to create a virtual laser, that would be in unshifted ZLS with C. Similarly we follow the laser B emission to a virtual position by moving it 2 ns towards C and D so that virtual laser B is now in unshifted ZLS with laser A. We have thus created a square configuration, indicated by the dashed line in Fig. 3b. From sub-lattice synchronization in a square geometry [20] we can easily determine that the revival of the correlation will take place at a time shift of ±32 ns around the central shifted ZLS peaks, corresponding to the total delay between pairs (A,B) and (C,D). Since our two virtual lasers lie along the space-timeline of the real laser configuration, the shifted correlations are the same for both the constructed square and real configurations.

The shifted ZLS between pairs (A,B) and (C,D) was confirmed in the simulations as shown in Fig. 3(c). The simulations also show the attenuated revival correlation peaks between diagonal pairs at a time shift of ±32 ns around the central shifted ZLS peaks. The cross correlation between adjacent lasers, exemplified by A and D in Fig. 3(c), is similar to a face-to-face configuration where attenuated peaks are observed at ±16 ns around the 4 ns shifted ZLS of laser D.

To complete the set of experimentally examined small motifs, we turn to the smallest motif which exhibits ZLS in the absence of self-feedback; an equilateral triangle, which was previously examined in the presence of self-feedback [21]. The experimental setup is similar to that of Fig. 1(a), where the mirror is replaced by a third laser and for the case of an equilateral triangle, the optical fiber is removed. The system is schematically presented in Fig. 4(b) and the delay times are tuned to be equal, $\tau_a=\tau_b=\tau_c$. Experimental results, with p=1.1, show ZLS between all three pairs of lasers, $\rho_{AB}$=93, $\rho_{BC}$=92, $\rho_{AC}$=91 at zero time shift, as shown in Fig. 4(c), consistent with simulations.

In conclusion, we experimentally demonstrate generalized ZLS in small motifs consisting of time-delay coupled chaotic SC lasers, without the limiting requirement of precisely matched delay times. For a quadrilateral geometry, sublattice synchronization is observed in agreement with numerical results. Depending on the specific geometric configuration, shifted and unshifted ZLS can be observed. Numerical results and theoretical arguments suggest the applicability of these experimental results to other configurations and larger motifs and open a route to new multi-user communication schemes which can be of great significance to modern communication networks.

tion of the chaotic intensities of all three pairs of lasers indicating ZLS.

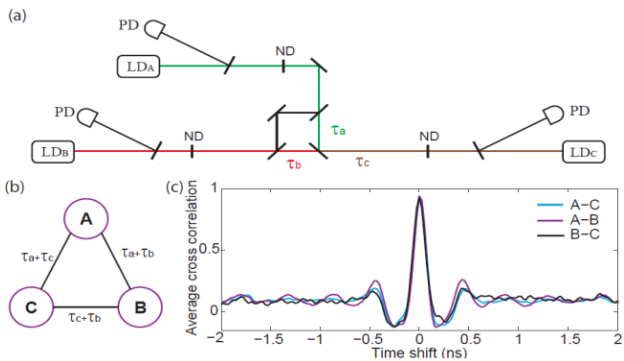

FIG. 4 (color online). (a) Experimental setup of three SC lasers in an equilateral triangle. Each laser is bidirectionally coupled to two adjacent lasers where $\tau_a=\tau_b=\tau_c$. (b) Schematic diagram of the experimental setup. (c) Shifted cross correlation